# Secondary Scintillation Yield in Pure Xenon

C.M.B. Monteiro[1], L.M.P. Fernandes[1], J.A.M. Lopes[1,2], L.C.C. Coelho[1], J.F.C.A. Veloso[1,3], J.M.F. dos Santos[1], K. Giboni[4], E. Aprile[4]

[1]GIAN, Physics Department, University of Coimbra, 3004-516 Coimbra, Portugal
[2]Instituto Superior de Engenharia de Coimbra, Apartado 4065, 3030-199 Coimbra, Portugal
[3]Physics Department, University of Aveiro, 3810-193 Aveiro, Portugal
[4]Physics Department and Columbia Astrophysics Laboratory, Columbia University, New York, NY 10027, USA

**ABSTRACT**

The xenon secondary scintillation yield was studied as a function of the electric field in the scintillation region, in a gas proportional scintillation counter operated at room temperature. A large area avalanche photodiode was used for the readout of the VUV secondary scintillation produced in the gas, together with the 5.9 keV x-rays directly absorbed in the photodiode. The latter was used as a reference for the determination of the number of charge carriers produced by the scintillation pulse and, thus, the number of VUV photons impinging the photodiode. A value of 140 photons/kV was obtained for the scintillation amplification parameter. The attained results are in good agreement with those predicted, for room temperature, by Monte Carlo simulation and Boltzmann calculations, as well as with those obtained for saturated xenon vapour, at cryogenic temperatures, and are about a factor of two higher than former results measured at room temperature.

**1. INTRODUCTION**

Xenon gas counters have been widely used as x-ray and low-energy γ-ray detectors [1,2]. More recently, xenon dual-phase detectors have been developed for dark matter search, namely for the detection of Weak Interacting Massive Particles [3-5]. In both cases, signal amplification in the gas is achieved by accelerating the electrons resulting from the radiation interaction to excite the gas atoms by electron impact, leading to the production of secondary scintillation, the so-called electroluminescence.

The electrons resulting from the radiation interaction in the gas are lead to a specific region inside the detector, the scintillation region, where the established electric field is such that electrons can acquire enough energy to excite the noble gas atoms through inelastic collisions. The most common type of collision that occurs is the elastic collision. Between two successive inelastic collisions, the electrons undergo a very large number of elastic collisions, above $10^4$ [6], while they gain enough energy from the electric field to excite the atoms. Nevertheless, the amount of energy lost by the electron in this large number of elastic collisions is small, given the very large difference in masses between electron and Xe atom. The excitation efficiency, i.e. the fraction of energy acquired from the electric field by the electron that is spent in exciting the xenon atoms, reaches values around 95% for reduced electric fields, E/p (the electric field divided by the gas pressure), of 4 kV cm$^{-1}$ bar$^{-1}$ [6,7]. On the other hand, this energy loss is not negligible for low reduced electric fields, as the number of elastic collisions increases significantly. The excitation efficiency presents a fast decrease for values of E/p below 2 kV cm$^{-1}$ bar$^{-1}$ [6,7], being zero below a characteristic E/p threshold. Below this threshold, electrons never acquire enough energy to excite the atoms of the medium.

The mechanisms of secondary scintillation production are well known [6,8-10]. For pure xenon, the wavelength of the emission depends on the gas pressure. Below 10 mbar, the emission is mainly atomic, with two peaks centred at ~130 and ~147 nm. For higher pressures, the formation of excited dimmers $Xe_2^*$ is favoured through three body collisions, and molecular emissions become increasingly important. These emissions are cantered at 147 nm (first continuum) and 172 nm (second continuum). The first and second continua correspond to the VUV radiative decay of the vibrationally $(Xe_2^*)^\upsilon$ excited and $(Xe_2^*)^{\upsilon=0}$ relaxed excimer states, respectively. Above 400 mbar, the second continuum is dominant and the secondary scintillation presents only a narrow peak, ~10 nm FWHM, centred at 172 nm [9]. The energy loss due to relaxation of the vibrationally excited dimmers is still a small fraction of the energy electrons acquire from the electric field and the overall scintillation efficiency reaches values of 80% for reduced electric fields of 4 kV cm$^{-1}$ bar$^{-1}$ [6,7].

Concerning the electroluminescence yield, defined as the number of secondary scintillation photons produced per drifting electron per unit path length, the data available in the literature are not in agreement. While the data obtained at room temperature using Monte Carlo simulation [7] and Boltzmann calculations [11] are in perfect agreement with each other [7], the values obtained experimentally [12-17] are much lower than the former and differ significantly from each other, see Fig.3 further on. On the other hand, the results presented for saturated gas at cryogenic temperatures [16,17], in equilibrium with the liquid phase, are in agreement with each other, as well as with the simulation results calculated for room temperature.

The electroluminescence yield, along with its dependence on the electric field, is an important parameter for detector simulation. Nevertheless, absolute measurements are difficult to perform and usually rely on comparison/calibration performed with experimental set-ups and/or settings other than those used to measure the secondary scintillation in itself, e.g. Ref. [17].

In this work, we apply a straightforward method that makes use of only one experimental set-up to carry out absolute electroluminescence yield measurements. A VUV-sensitive large area avalanche photodiode (LAAPD) is used to detect, simultaneously, the secondary scintillation of a xenon gas proportional scintillation counter (GPSC) and original x-rays. The x-rays are used as reference for determining the absolute number of VUV-photons impinging the LAAPD. This method has been largely employed to measure the primary scintillation yield of inorganic crystals, e.g. Ref. [18] and references therein, and the primary scintillation yield produced by alpha particles in noble gases [19,20]. The electroluminescence yield is measured as a function of electric field in the scintillation region and is compared with the other experimental and calculated results found in the literature.

## 2. EXPERIMENTAL SET-UP

In this work we have used a xenon-filled GPSC instrumented with a large area APD [21]. The GPSC is schematically depicted in Fig. 1 and was already used in Refs. [22,23]. It is a driftless prototype with a 1.1-cm deep scintillation region. The APD is positioned just below the electron-collection grid, G (80 µm in diameter stainless steel wire with 900-µm spacing, having an optical transparency $T$ of 84%). The radiation window is maintained at negative high-voltage, -HV, while grid G, APD and detector enclosures are maintained at ground potential. The electric field in the scintillation region is defined by -HV. The radiation window is a 12.5-µm thick, 10 mm in diameter aluminized Mylar film. A Macor ceramic is used between the detector body and the radiation window holder, for electrical insulation. Both Mylar window and Macor are glued to the stainless steel by means of a low vapour pressure epoxy (TRA-CON 2116). The detector upper and lower parts and the LAAPD enclosure are vacuum-tight by compression of indium gaskets.

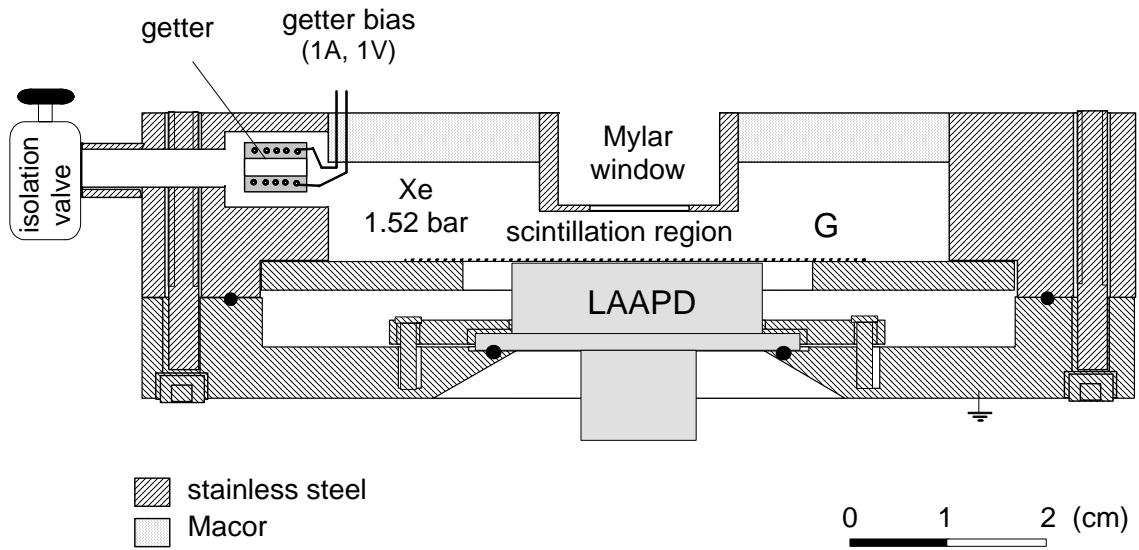

Fig.1 – Schematic of the driftless GPSC used in the present work.

The detector is pumped down to pressures in the $10^{-6}$ mbar range, and filled with high purity xenon (Air Liquide, N45) at a pressure of 1.52 bar. After being sealed, the gas purity in the detector is maintained by a small non-evaporable getter (SAES ST172/HI/7-6/150C), which is placed inside the xenon envelope and kept at a temperature of about 200ºC. These simple, high-vacuum detector assembling techniques have been developed and used in our group for more than one decade and have proved to be capable of keeping the xenon in sealed detectors at high purity levels, with high scintillation efficiency. This produces high performance detectors capable to deliver the best energy resolutions obtained with GPSCs [1].

For this study, we have used a 1-mm collimated 5.9-keV x-ray beam from a $^{55}$Fe x-ray source filtered with a Cr-film to remove the 6.4-keV Mn K$_\beta$-fluorescence line. X-rays entering the radiation window are absorbed in the xenon gas by photoelectric effect, generating a primary electron cloud. The electrons are accelerated in the electric field established by -HV. Each primary electron produces a large number of VUV photons. Proportionality between incident x-ray energy, number of primary electrons and number of scintillation photons is maintained throughout the process.

The scintillation photons absorbed in the sensitive area of the APD produce electron–hole pairs in the silicon, which are multiplied by the avalanche process. Concurrent with the acquisition of the scintillation signals resulting from the absorption of x-rays in xenon, a transmitted fraction of the x-rays is detected directly by the APD. The number of electron–hole pairs produced through direct absorption of the x-rays in the APD is determined from the energy of the x-ray and the w-value in silicon, i.e. the mean energy required to produce a pair of charge carriers. The APD reverse bias voltage determines the multiplication gain in the avalanche process.

In the driftless design, the primary scintillation produced by the x-ray interaction is detected together with the secondary scintillation. However, the number of primary photons is more than two orders of magnitude lower than the number of photons resulting from secondary scintillation.

For each 5.9-keV x-ray interaction in the driftless detector, the total number of scintillation photons produced by the primary electron cloud will depend on how deep into the scintillation region the x-ray is absorbed. Nevertheless, the average number of VUV photons is well defined, as the respective pulse-height distribution has a Gaussian shape with a tail towards the low energy region. Although the driftless GPSC results in degraded energy resolution for scintillation events, it allows a higher transmission of the 5.9-keV x-rays through xenon and, therefore, more direct x-ray interactions in the APD, convenient for this study. In our case, approximately 0.2% of the 5.9-keV x-rays are transmitted through 1.1 cm of xenon.

The charge pulses collected in the APD are integrated in a 1.5 VpC$^{-1}$ charge-sensitive preamplifier (Canberra 2004), followed by linear amplification (Hewlett Packard 5582A) with a 2-µs shaping time. Pulse-height analysis is performed with a 1024-channel analyser (Nucleus PCA-II). The peaks in the pulse-height distribution are fit to a Gaussian function superimposed on a linear background. The pulse amplitude for each type of event is determined from the centroid of the fitted Gaussian.

## 3. METHOD

Figure 2 depicts a typical pulse height distribution obtained with the driftless GPSC instrumented with an LAAPD for scintillation readout. The salient features of the pulse-height distribution include, not only the 5.9-keV x-ray full-energy peak from absorption in the xenon GPSC, but also the xenon L$_\alpha$- and L$_\beta$-escape peaks from 5.9-keV x-ray absorption in the xenon GPSC, the 5.9-keV x-ray peak from direct absorption in the APD, the 4.1 and 4.8 keV xenon L$_\alpha$ and L$_\beta$ fluorescence peaks from absorption in the APD, and the system electronic noise. While the amplitude of the scintillation peaks depends on both scintillation region and LAAPD biasing, the amplitude of the events resulting from direct x-ray interaction in the LAAPD depends only on the LAAPD biasing. The latter events are present even for a null or reverse electric field in the scintillation region. On the other hand, only the peak resulting from direct 5.9-keV interaction in the LAAPD is present when the detector is vacuum-pumped. Therefore, the pulse-height distributions enable a direct comparison between the pulse amplitudes resulting from the 5.9-keV x-ray full-absorption in the gas, i.e. from the xenon scintillation, and direct absorption in the APD. This allows a direct quantification of the VUV-photons impinging the LAAPD, given the quantum efficiency of the device.

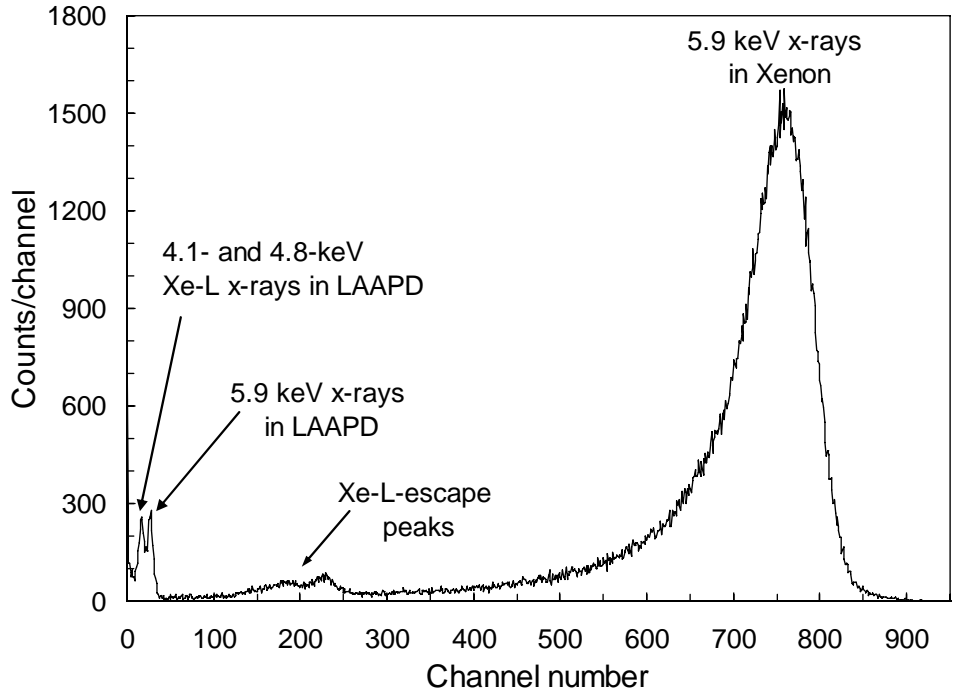

Fig.2 – Pulse-height distribution from a xenon driftless GPSC instrumented with a large-area APD, for 5.9-keV X-rays. An E/p of 4.1 kVcm$^{-1}$bar$^{-1}$ was used in the scintillation region.

According to the manufacturer, the LAAPD manufacturing technology is well established, and quite good reproducibility is obtained. Thus, it is expected that the observed behaviour for individual LAAPDs is representative for any of these devices [24]. A value of $QE = 1.1$ for the number of charge carriers produced in the photodiode per incident 172-nm VUV photon was provided by the manufacturer [25] for the LAAPDs we acquired [26]. We note that, at the moment, the manufacturer provides different values for the quantum efficiency of the LAAPDs presently manufactured which, in the VUV region [21], are somewhat higher than formerly. We assume an uncertainty of ~ 0.1 for the LAAPD quantum efficiency. This uncertainty is the major source of error in our measurements.

For a 4.1 kVcm$^{-1}$bar$^{-1}$ reduced electric field, the ratio between the pulse amplitudes resulting from the 5.9 keV x-ray full-absorption in the gas and absorbed in the APD, as obtained from the corresponding pulse-height distribution, is $A_{UV} / A_{XR} = 19.9 \pm 0.1$ for low LAAPD gains, where gain non-linearity in the photodiode is less than 1% [22]. As the w-value in silicon is 3.62 eV [27], the average number of free electrons produced in the LAAPD by full absorption of the 5.9-keV x-rays is

$$N_{XR} = \frac{5895 \text{ eV}}{3.62 \text{ eV}} \cong 1.63 \times 10^3 \text{ electrons} \qquad (1)$$

Thus, the average number of VUV photons impinging the photodiode for the scintillation pulses due to the 5.9-keV x-ray full-absorption in the gas is

$$N_{UV,APD} = \frac{A_{UV}}{A_{XR}} \times \frac{N_{XR}}{QE} \cong 2.95 \times 10^4 \text{ photons} \qquad (2)$$

The average solid angle $\Omega$ subtended by the photosensor active area for the primary electron path has been computed accurately through a simple Monte Carlo simulation program [28]. A value of

$$\Omega_{rel} = \frac{\Omega}{4\pi} \cong 0.202 \qquad (3)$$

was obtained for the present geometry. In this way, the total number of VUV photons produced by the full absorption of 5.9-keV x-rays in the detector is

$$N_{UV,total} = \frac{N_{UV,APD}}{\Omega_{rel} \times T} \cong 1.74 \times 10^5 \text{ photons} \qquad (4)$$

where $T$ is the grid optical transparency.

In addition, the average number of primary electrons produced in xenon by the full absorption of the 5.9-keV x-rays is

$$N_e = \frac{5895 \text{ eV}}{22.4 \text{ eV}} \cong 263 \text{ electrons} \qquad (5)$$

considering a w-value for xenon of 22.4 eV [29]. As the average absorption depth of 5.9 keV x-rays in the detector is given by

$$d_{av} = \frac{\int_0^{1.1} x \, e^{-\lambda x} dx}{\int_0^{1.1} e^{-\lambda x} dx} \cong 0.17 \text{ cm} \qquad (6)$$

where $\lambda = 3.90$ cm$^{-1}$ is the linear attenuation coefficient [30]. The average distance the primary electron cloud drifts in the detector is 0.93 cm.

In this way, the reduced scintillation yield, $Y/p$, determined for 4.1 kVcm$^{-1}$bar$^{-1}$, is

$$\frac{Y}{p} = \frac{N_{UV,total}}{N_e \times d_{av} \times p} = 466 \qquad (7)$$

photons per electron per cm of path and per bar.

## 4. EXPERIMENTAL RESULTS AND DISCUSSION

In Fig.3 we depict the reduced electroluminescence yield, Y/N, i.e. the electroluminescence yield divided by the number density of the gas as a function of reduced electric field, E/N, in the scintillation region. Other results presented in the literature are also depicted for comparison. The reduced electroluminescence yield shows an approximately linear variation with the reduced electric field which can be represented by

$$\frac{Y}{N}\left(10^{-17} \text{ photons electron}^{-1} \text{ cm}^2 \text{ atom}^{-1}\right) = 0.140\frac{E}{N} - 0.474 \qquad (8)$$

where $E/N$ is given in Td ($10^{-17}$ V cm$^2$ atom$^{-1}$).

It can also be expressed by:

$$\frac{Y}{p}\left(\text{photons electron}^{-1} \text{ cm}^{-1} \text{ bar}^{-1}\right) = 140\frac{E}{p} - 116 \qquad (9)$$

where $E/p$ is given in kV cm$^{-1}$ bar$^{-1}$.

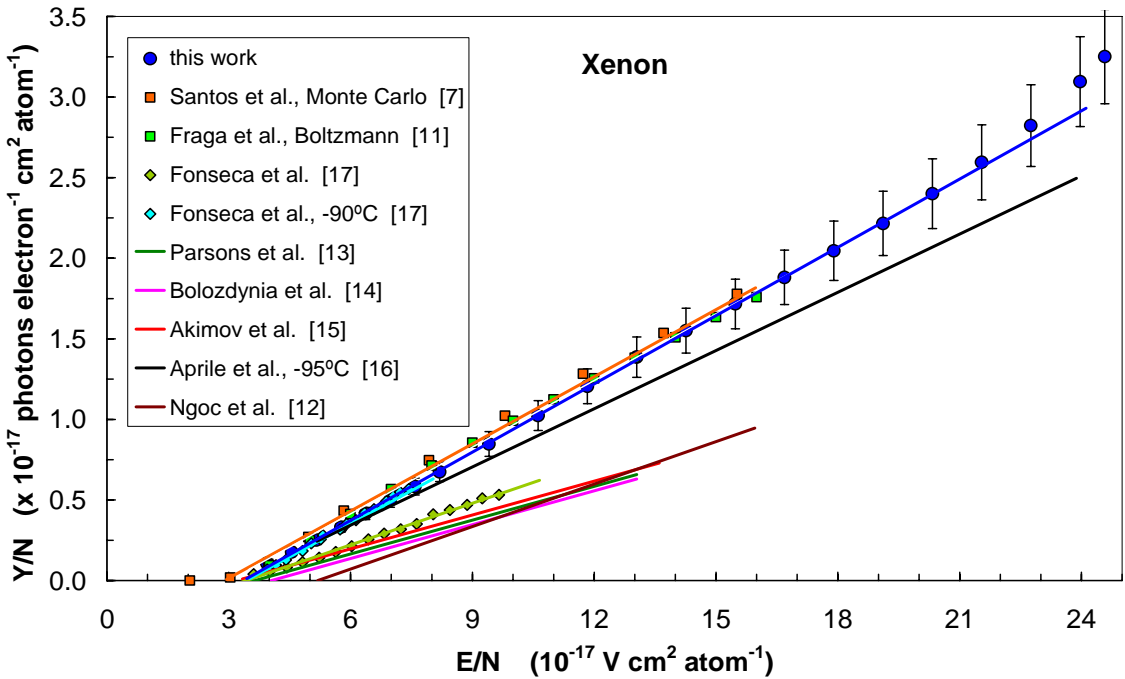

Fig.3 – Xenon reduced electroluminescence yield as a function of reduced electric field for this work, as well as for the different data reported in the literature.

Our results are in good agreement with those obtained by Monte Carlo and Boltzmann simulations for room temperature.

Since the early years of study and development of GPSCs, it was well established that the reduced scintillation yield presents an approximately linear dependence on the reduced electric field in the scintillation region, presenting a scintillation threshold at an $E/p$ value of about 1.3 kVcm$^{-1}$bar$^{-1}$ (e.g. Ref. [31] and references therein), as exhibited in Fig.3. However, the scintillation amplification parameter - the number of photons produced per drifting electron and per volt, i.e. the slope of the linear dependence - presents many different values in the literature and is not well established yet. Above a reduced field of 8 kVcm$^{-1}$bar$^{-1}$, the xenon ionisation threshold, the reduced scintillation yield variation departs from the linear behaviour, reflecting the exponential growth in the number of electrons present in the scintillation region, since secondary electrons also produce electroluminescence, while Y/N is calculated per primary electron. Favata et al. [32] reported a detailed study of the electroluminescence yield as a function of reduced electric field and compiled the different studies on electroluminescence yield published up to then (1990), concluding that the reduced electroluminescence yield is pressure-independent. Fonseca et al. [17] have shown that the scintillation yield does not depend on gas temperature, in the range from 20 downto -88ºC. On the other hand, for -90ºC and at a constant gas pressure, the scintillation amplification factor near the saturation point varies significantly, depending on the amount of xenon present in the liquid phase.

Table I summarizes the different values found in the literature for the scintillation amplification parameter, and the respective experimental conditions concerning pressure and temperature. The value of 70 photons/kV, obtained by Parsons et al. (1989) [13] and confirmed by Bolozdynya et al. (1997) [14] and Akimov et al. (1997) [15] is a factor of two lower than that obtained by Monte Carlo simulation and Boltzmann calculation. In 2003, Aprile et al. [16] have estimated a value of 120 photons/kV from their experimental measurements in saturated xenon vapour, a value in much better agreement with the simulation/calculation. The value obtained by Fonseca et al. [17] for the amplification parameter in saturated xenon vapour, at -90ºC, is in very good agreement with that obtained by simulation/calculation for room temperature, but their value obtained at room temperature is only ~90 photons/kV, a similar value to what had been obtained by Ngoc et al. (1980) [12].

Table I – Xenon secondary scintillation amplification parameter, reduced electroluminescence yield linear trends and experimental conditions of pressure and temperature for this work, as well as for the different data reported in the literature.

| Work | Amplification parameter (photons/kV) | linear trend | | Pressure | Temperature |
|---|---|---|---|---|---|
| | | Density units* | Pressure units** | | |
| Present work | 140 | $Y/N = 0.140\ E/N - 0.474$ | $Y/p = 140\ E/p - 116$ | 1.52 bar | 20ºC |
| Santos et al. (MC) [7] | 139 | $Y/N = 0.139\ E/N - 0.402$ | $Y/p = 139\ E/p - 100$ | 1 atm | 20ºC |
| Fraga et al. (Boltzmann) [11] | 138 | $Y/N = 0.138\ E/N - 0.413$ | $Y/p = 138\ E/p - 102$ | 1 atm | 20ºC |
| Fonseca et al. [17] | 137 | $Y/N = 0.137\ E/N - 0.470$ | $Y/p = 137\ E/p - 125$ | 2 bar | -90ºC |
| Aprile et al. [16] | 120 | $Y/N = 0.120\ E/N - 0.378$ | $Y/p = 120\ E/p - 154$ | 2 atm | -95ºC |
| Ngoc et al. [12] | 88 | $Y/N = 0.088\ E/N - 0.479$ | $Y/p = 88\ E/p - 113$ | 2 - 10 atm | 20ºC |
| Fonseca et al. [17] | 86 | $Y/N = 0.086\ E/N - 0.296$ | $Y/p = 86\ E/p - 73$ | 2 bar | 20ºC |
| Parsons et al. [13] | 70 | $Y/N = 0.070\ E/N - 0.255$ | $Y/p = 70\ E/p - 63$ | 5 atm | 20ºC |
| Bolozdynya et al. [14] | 70 | $Y/N = 0.070\ E/N - 0.283$ | $Y/p = 70\ E/p - 70$ | - | 20ºC |
| Akimov et al. [15] | 70 | $Y/N = 0.070\ E/N - 0.224$ | $Y/p = 70\ E/p - 56$ | 2, 4, 8 atm | 20ºC |

\* $E/N$ in Td ($10^{-17}$ V cm$^2$ atom$^{-1}$)

\*\* $E/p$ in kV cm$^{-1}$ bar$^{-1}$

Our measurements have shown, for the first time that, even for room temperature, the scintillation amplification parameter can be as high as those predicted by Monte Carlo simulation and/or Boltzmann analysis and those experimentally obtained for saturated xenon vapour at cryogenic temperatures. We attribute the differences in the experimental values obtained at room temperature to different levels of gas purity achieved in each experimental set-up. Electron collisions with molecular impurities lead to energy losses through excitation of rotational and vibrational molecular states, which de-excite without emitting scintillation. Thus, higher impurity content will result in less efficient energy transfer from the electric field to photons, leading to lower scintillation amplification values.

## 4. CONCLUSIONS

We have studied the reduced electroluminescence yield of pure xenon at room temperature and compared it with the other results reported in the literature. The experimental measurements were performed with a gas proportional scintillation counter (GPSC) instrumented with a large area avalanche photodiode for the VUV secondary scintillation readout. X-rays with energy of 5.9 keV were used to induce the secondary scintillation in the GPSC or partially to interact in the photodiode. The direct interactions are used as a reference for the determination of the number of charge carriers produced by the scintillation pulse and, thus, the number of VUV photons impinging the photodiode, given its quantum efficiency.

A scintillation amplification parameter, i.e., number of photons produced per drifting electron and per volt, of 140 photons/kV, was measured. The results are in good agreement with those predicted by Monte Carlo simulation and Boltzmann calculation for room temperature and also with those observed for saturated xenon vapour at cryogenic temperatures. Our result is about a factor of two higher than earlier results measured at room temperature. Differences in gas purity during the experimental measurements may be one of the factors responsible for the different experimental results obtained at room temperature.

## ACKNOWLEDGEMENTS

This work received support from FEDER and POCI2010 programs, through FCT (Fundação para a Ciência e Tecnologia) project POCI/FIS/60534/2004. C.M.B. Monteiro acknowledges to FCT grant BD/35569/2005.

## REFERENCES

[1] J.M.F. Dos Santos, J.A.M. Lopes, J.F.C.A. Veloso, P.C.P.S. Simões, T.H.V.T. Dias, F.P. Santos, P.J.B.M. Rachinhas, L.F. Requicha Ferreira, C.A.N. Conde, Development of portable gas proportional scintillation counters for x-ray spectrometry, X-Ray Spectrom. 30 (2001) 373.

[2] C.A.N. Conde, Gas Proportional Scintillation Counters for X-ray Spectrometry, in X-Ray Spectrometry: Recent Technological Advances, Eds. K. Tsuji, J. Injuk and R. van Grieken, John Wiley & Sons (2004), ISBN: 0-471-48640-X.

[3] E.Aprile et al., The XENON Dark Matter Experiment, New Astronomy Reviews, 49 (2005) 289-295

[4] H.M. Araújo et al. The ZEPLIN-III dark matter detector: performance study using an end-to-end simulation tool, Astroparticle Physics 26 (2006) 140.

[5] H. Wang, Xenon as a detector for dark matter search, Phys. Reports 307 (1998) 263.

[6] T.H.V.T. Dias, F.P. Santos, A.D. Stauffer, C.A.N. Conde, Monte Carlo simulation of x-ray absorption and electron drift in gaseous xenon, Phys. Rev. A 48 (1993) 2887.

[7] F.P. Santos, T.H.V.T. Dias, A.D. Stauffer, C.A.N. Conde, Three dimensional Monte Carlo calculation of the VUV electroluminescence and other electron transport parameters in xenon, J. Phys. D: Appl. Phys. 27 (1994) 42.

[8] M. Suzuki, S. Kubota, Mechanism of proportional scintillation in argon, krypton and xenon, Nucl. Instrum. Meth. 164 (1979) 197.

[9] M.S.C.P. Leite, Radioluminescence of rare gases, Portugal. Phys. 11 (1980) 53.

[10] T. Takahashi, S. Himi, J. Ruan, S. Kubota, Emission spectra from Ar-Xe, Ar-Kr, Ar-N2, Ar-CH4, Ar-CO2 and Xe-N2 gas scintillation proportional counters, Nucl. Instrum. Meth. 205 (1983) 591.

[11] M.M.F.R. Fraga, C.M. Ferreira, J. Loureiro, M.S.C.P. Leite, presented at the 1990 ESCAMPIG, 28-31 Aug., Orleans, France.

[12] H.N. Ngoc, J. Jeanjean, H. Itoh, G. Charpak, A xenon high-pressure proportional scintillation camera for x and γ-ray imaging, Nucl. Instrum. Meth. 172 (1980) 603.

[13] A. Parsons, B. Sadoulet, S. Weiss, T. Edberg, J. Wilkerson, G. Smith, R.P. Lin, K. Hurley, High pressure gas scintillation drift chambers with wave shifter fiber readout, IEEE Trans. Nucl. Sci. 36 (1989) 931.

[14] A. Bolozdynya, V. Egorov, A. Koutchenkov, G. Safronov, G. Smirnov, S. Medved, V. Morgunov, A high-pressure xenon self-triggered scintillation drift chamber with 3D sensitivity in the range of 20-140 keV deposited energy, Nucl. Instrum. Meth. A 385 (1997) 225.

[15] D.Y. Akimov, A.A. Burenkov, D.L. Churakov, V.F. Kuzichev, V.L. Morgunov, G.N. Smirnov, V.N. Solovov, Development of high pressure Xe scintillation proportional counter for experiments in "low-background" physics, arXiv:hep-ex/9703011 v1 20 Mar 1997.

[16] E. Aprile, K.L. Giboni, P. Majewski, K. Ni, M. Yamashita, Proportional Light in a Dual-Phase Xenon Chamber, IEEE Trans. Nucl. Sci. 51 (2004) 1986; presented at the IEEE Nucl. Sci. Symp., November 2003, Seattle, USA.

[17] A.C. Fonseca, R. Meleiro, V. Chepel, A. Pereira, V. Solovov and M.I. Lopes, Study of Secondary Scintillation in Xenon Vapour, 2004 IEEE Nucl. Sci. Symp. Conference Record (2005).

[18] M. Moszynski, M. Szawlowski, M. Kapusta, M. Balcerzyk, Large area avalanche photodiodes in scintillation and X-rays detection, Nucl. Instrum. Meth. A 485 (2002) 504-521.

[19] K. Saito, H. Tawara, T. Sanami, E. Shibamura, S. Sasaki, Absolute numbers of scintillation photons emitted by alpha particles in rare gases, IEEE Trans. Nucl. Sci. 49 (2002) 1674-1680.

[20] K. Saito, S. Sasaki, H. Tawara, T. Sanami, E. Shibamura, Simultaneous measurements of absolute numbers of electrons and scintillation photons produced by 5.49 MeV alpha particles in rare gases, IEEE Trans. Nucl.Sci. 50 (2003) 2452-2459.

[21] Deep-UV series, Advanced Photonix Inc., http://www.advancedphotonix.com/

[22] L.M.P. Fernandes, J.A.M. Lopes, C.M.B. Monteiro, J.M.F. Dos Santos, C.A.N. Conde, Non-linear behavior of large-area avalanche photodiodes, Nucl. Instrum. Meth. A 478 (2002) 395.

[23] J.F.C.Veloso, J.A.M. Lopes, C.A.N.Conde, L.M.P. Fernandes, E.D.C. Freitas, O. Huot, P. Knowles, F: Kottmann, F. Mulhauser, J.M.F. dos Santos and D. Taqqu, Gas proportional scintillation counters for the p Lamb-shift experiment, IEEE Trans. Nucl. Sci. 49 (2002) 899.

[24] Szawlowski, Advanced Photonix Inc., private communication (2002).

[25] B. Zhou, M. Szawlowski, An explanation on the APD spectral quantum efficiency in the deep UV range, Interoffice Memo, Advanced Photonix Inc., 1240 Avenida Acaso, Camarillo, CA 93012, EUA, 1999.

[26] J.A.M. Lopes, J.M.F. Dos Santos, R.E. Morgado, C.A.N. Conde, A xenon gas proportional scintillation counter with a UV-sensitive, large-area avalanche photodiode, IEEE Trans. Nucl. Sci., 48 (2001) 312-319.

[27] G.F. Knoll, Radiation Detection and Measurement, 3$^{rd}$ Edition, Wiley, New York, 2000.

[28] J.M.F. dos Santos, A.C.S.M. Bento, C.A.N. Conde, The dependence of the energy resolution of gas proportional scintillation counters on the scintillation region to photomultiplier distance, IEEE Trans. Nucl. Sci. 39 (1992) 541.

[29] T.H.V.T. Dias, J.M.F. dos Santos, P.J.B.M. Rachinhas, F.P. Santos, C.A.N. Conde, A.D. Stauffer, Full-energy absorption of x-ray energies near the Xe L-and K-photoionization thresholds in xenon gas detectors: Simulation and experimental results, J. Appl. Phys. 82 (1997) 2742.

[30] http://physics.nist.gov/PhysRefData/Xcom/html/xcom1.html

[31] M.A. Feio, A.J.P.L. Policarpo, M.A.F. Alves, Jap. J. Appl. Phys. 8 (1982) 1184.

[32] F. Favata, A. Smith, M. Badvaz, T. Kowalski, Light yield as a function of gas pressure and electric field in gas scintillation proportional counters, Nucl. Instrum. Meth. A 294 (1990) 595.